\documentstyle[12pt]{article}

\newfont{\tenof}{msym10}
\newcommand{\R}{\mbox{\tenof R}}
\newcommand{\F}{\Phi_{\sigma}}
\newcommand{\f}[1]{\varphi^{[f]}_{#1}}
\newcommand{\V}[2]{V^{[f]}_{#1}(#2)}
\newcommand{\ts}[1]{\tau^{(#1)}_{\sigma}}
\newcommand{\ls}[1]{\lambda_{\sigma(#1)}}
\newcommand{\sumij}{\sum_{\scriptstyle i,j \atop \scriptstyle i\ne j}}
\newcommand{\sumijk}{\sum_{\scriptstyle i,j,k \atop \scriptstyle i\ne j\ne
k}}
\newcommand{\ordsum}{\sum_{\scriptstyle i,j,k \atop \scriptstyle i<j<k}}
\newcommand{\csch}{\mathop{\rm csch}\nolimits}

\begin{document}

\baselineskip=22pt plus 1pt minus 1pt
%
%--------------------------------------------------------------------------
%
\noindent{\Large \bf Knizhnik-Zamolodchikov equations and}

\noindent{\Large \bf the Calogero-Sutherland-Moser integrable models}

\noindent{\Large \bf with exchange terms}

\bigskip\bigskip
\noindent
{\bf C\ Quesne\footnote{Directeur de recherches FNRS}$^,$\footnote{E-mail
address: cquesne@ulb.ac.be}}

\bigskip
\noindent\rm Physique Nucl\'eaire Th\'eorique et Physique Math\'ematique,
Universit\'e Libre de Bruxelles, Campus de la Plaine CP229, Boulevard du
Triomphe, B-1050 Brussels, Belgium

\vspace{8cm}
\noindent Short title: {\sl Knizhnik-Zamolodchikov equations}

\bigskip
\noindent PACS number(s): 03.65, 11.30

\vfill\eject
%
%==================================================================
%
\noindent{\bf Abstract}

It is shown that from some solutions of generalized Knizhnik-Zamolodchikov
equations one can construct eigenfunctions of the Calogero-Sutherland-Moser
Hamiltonians with exchange terms, which are characterized by any given
permutational symmetry under particle exchange. This generalizes some
results previously derived by Matsuo and Cherednik for the ordinary
Calogero-Sutherland-Moser Hamiltonians.
\vfill\eject
%
%===========================================================================
%
\section{Introduction}
Recently, much attention has been paid to the Calogero-Sutherland-Moser (CSM)
integrable systems~\cite{1}, \cite{2},~\cite{3} both in field-theoretical and
in condensed-matter contexts. They are indeed relevant to several apparently
disparate physical problems, such as fractional statistics and
anyons~\cite{4}, spin chain models~\cite{5}, soliton wave
propagation~\cite{6}, two-dimensional nonperturbative quantum gravity and
string theory~\cite{7}, and two-dimensional QCD~\cite{8}.\par
%
%--------------------------------------------------------------------------
%
Such one-dimensional integrable systems consist of $N$ nonrelativistic
particles interacting through two-body potentials of the inverse square type
and its generalizations, and are related to root systems of ${\cal A}_{N-1}$
algebras~\cite{9}. Their spectra and wave functions can be obtained by
simultaneously diagonalizing a set of $N$ commuting first-order differential
operators, first considered by Dunkl in the mathematical
literature~\cite{10}, and later rediscovered by Polychronakos~\cite{11} and
Brink {\it et al\/}~\cite{12}. The use of Dunkl operators leads to
Hamiltonians with exchange terms, related to the spin generalizations of the
CSM models~\cite{13}.\par
%
%---------------------------------------------------------------------------
%
Dunkl operators are rather similar~\cite{14} to the differential operators
of the Knizhnik-Zamolodchikov (KZ) equations, which first appeared in
conformal field theory~\cite{15}. Matsuo~\cite{16} and Cherednik~\cite{17}
proved that from some solutions of the KZ~equations, one can construct wave
functions for the (ordinary) CSM~systems. Such relations between
KZ~equations and CSM~systems were then exploited by Felder and
Veselov~\cite{18} to provide a natural interpretation for the shift
operators of the latter.\par
%
%---------------------------------------------------------------------------
%
The purpose of the present paper is to extend Matsuo and Cherednik results to
some CSM models with exchange terms. In the following section, we review
generalized KZ~equations. Then, in section~3, we establish new links
between some of their solutions and wave functions of corresponding CSM~models
with exchange terms. Finally, section~4 contains the conclusion.\par
%
%===========================================================================
%
\section{Generalized Knizhnik-Zamolodchikov equations}
Let us consider a system of $N$ first-order partial differential equations
of the type
\begin{equation}
  \partial_i\Phi = \left(\sum_{j\ne i} \left(f_{ij}(x_i-x_j) P^{(ij)} +
    c\, T^{(ij)}\right) + \lambda^{(i)}\right) \Phi \qquad i=1,2,\ldots,N
    \label{eq:1}
\end{equation}
where $\Phi = \Phi(x_1,x_2,\ldots,x_N)$ takes values in the tensor product
$V\otimes V\otimes\cdots\otimes V = V^{\otimes N}$ of some $N$-dimensional
vector space $V$, $f_{ij}(x_i-x_j)$ is a function of $x_i-x_j$, and $c$ is a
complex parameter. Equation~(\ref{eq:1}) also contains three operators
$P^{(ij)}$, $T^{(ij)}$, and $\lambda^{(i)}$, defined as in ref.~\cite{18},
i.e., $\lambda^{(i)}$ is the operator in $V^{\otimes N}$ acting on the
$i^{\rm th}$ factor as the diagonal matrix $\lambda = \hbox{\rm
diag}(\lambda_1, \ldots, \lambda_N)$ and identically on all other factors,
$P$ is the permutation: $P(a\otimes b) = b\otimes a$, $T$ is the following
operator on $V\otimes V$:
\begin{equation}
  T = \sum_{k>l} \left(E_{kl}\otimes E_{lk} - E_{lk}\otimes E_{kl}\right)
    \label{eq:2}
\end{equation}
where $E_{kl}$ denotes the $N\times N$ matrix with entry~1 in row~$k$ and
column~$l$ and zeroes everywhere else, and $P^{(ij)}$ and $T^{(ij)}$ are the
corresponding operators in $V^{\otimes N}$ acting only on the $i^{\rm th}$
and $j^{\rm th}$ factors. When $\lambda_i = c = 0$, and $f_{ij}(x_i-x_j) = k
(x_i-x_j)^{-1}$, the set of equations~(\ref{eq:1}) coincides with that derived
by Knizhnik and Zamolodchikov in conformal field theory~\cite{15}. We shall
therefore refer to~(\ref{eq:1}) as generalized KZ~equations.\par
%
%---------------------------------------------------------------------------
%
Let us consider the case where $\Phi$ has the form
\begin{equation}
  \Phi = \sum_{\sigma\in S_N} \F e_{\sigma} \qquad e_{\sigma}
    = e_{\sigma(1)} \otimes e_{\sigma(2)} \otimes \cdots \otimes
    e_{\sigma(N)} \label{eq:3}
\end{equation}
where $S_N$ is the symmetric group, and $e_k$ denotes a column vector with
entry~1 in row~$k$ and zeroes everywhere else. The operators $P^{(ij)}$,
$T^{(ij)}$, and~$\lambda^{(i)}$ transform the components $\F$ into
$\Phi_{\sigma \circ p_{ij}}$, $\ts{ij} \Phi_{\sigma  \circ p_{ij}}$, and
$\ls{i} \F$, respectively, where $p_{ij}\in S_N$ is the transposition of~$i$
and~$j$, and $\ts{ij} \equiv \hbox{\rm sgn}\bigl(\sigma(i) -
\sigma(j)\bigr)$ satisfies the relations
\begin{eqnarray}
  & & \ts{ij} = - \tau^{(ij)}_{\sigma \circ p_{ij}} = - \ts{ji} \qquad
      \tau^{(ik)}_{\sigma \circ p_{ij}} = \ts{jk} \qquad
      \tau^{(kl)}_{\sigma \circ p_{ij}} = \ts{kl} \nonumber \\
  & & \ts{ij} \ts{ik} + \ts{jk} \ts{ji} + \ts{ki} \ts{kj} = 1
      \label{eq:4}
\end{eqnarray}
for any $i\ne j \ne k \ne l$. Hence, for such functions $\Phi$,
equation~(\ref{eq:1}) is equivalent to the set of equations
\begin{equation}
  \partial_i\F = \sum_{j\ne i} \left(f_{ij}(x_i-x_j) + c\, \ts{ij}\right)
  \Phi_{\sigma \circ p_{ij}} + \ls{i} \F \qquad i=1,2,\ldots,N \label{eq:5}
\end{equation}
where $\sigma$ is an arbitrary permutation of $S_N$.\par
%
%--------------------------------------------------------------------------
%
The integrability conditions of~(\ref{eq:5}), i.e., $\partial_j \partial_i
\Phi_{\sigma} = \partial_i \partial_j \Phi_{\sigma}$ for any $i$, $j=1$,~2,
$\ldots$,~$N$, and any $\sigma \in S_N$, are satisfied if and only if
\begin{equation}
  f_{ij}(x_i-x_j) = - f_{ji}(x_j-x_i) \label{eq:6}
\end{equation}
\begin{equation}
  f_{ij}(x_i-x_j) f_{jk}(x_j-x_k) + f_{jk}(x_j-x_k) f_{ki}(x_k-x_i) +
  f_{ki}(x_k-x_i) f_{ij}(x_i-x_j) = - c^2 \label{eq:7}
\end{equation}
for any $i$,~$j$, $k=1$,~2, $\ldots$,~$N$, such that $i\ne j\ne k$. By
taking~(\ref{eq:6}) into account, eq.~(\ref{eq:7}) can be rewritten as
\begin{equation}
  f_{ij}(u) f_{jk}(v) - f_{ik}(u+v) \left[f_{ij}(u) + f_{jk}(v)\right] =
  - c^2. \label{eq:8}
\end{equation}
It is enough to consider the latter for $1\le i<j<k\le N$, since the
relations corresponding to different orderings of $i$, $j$,~$k$ directly
follow from them.\par
%
%---------------------------------------------------------------------------
%
Equation~(\ref{eq:8}) looks like a functional equation first considered by
Sutherland~\cite{2}, and solved by Calogero~\cite{19} through a small-$x$
expansion. By using a similar procedure, all the solutions of~(\ref{eq:8})
that are odd and meromorphic in a neighbourhood of the origin can be easily
derived. Denote by $F(u)$ and $G(u)$ the functions
\begin{equation}
  F(u) = \cases{
      k \omega \coth \omega u & if $c^2 = k^2\omega^2 >0$ \cr
      \noalign{\smallskip}
      k/u                     & if $c^2 = 0$ \cr
      \noalign{\smallskip}
      k \omega \cot \omega u  & if $c^2 = -k^2\omega^2 <0$ \cr}
  \label{eq:9}
\end{equation}
and
\begin{equation}
  G(u) = \cases{
      k \omega \tanh \omega u  & if $c^2 = k^2\omega^2 >0$ \cr
      \noalign{\smallskip}
      - k \omega \tan \omega u & if $c^2 = -k^2\omega^2 <0$ \cr}
  \label{eq:10}
\end{equation}
where $\omega \in \R^+$. For any $N\ge3$ and $c^2\ne0$, one finds that
equation~(\ref{eq:8}) has two and only two types of odd, meromorphic solutions,
namely
\begin{equation}
  f_{ij}(u) = f_{ji}(u) = F(u) \qquad 1\le i<j\le N \label{eq:11}
\end{equation}
and
\begin{equation}
  f_{ij}(u) = f_{ji}(u) = \cases{
      F(u) & if $1\le i<j\le N_1$ or $N_1+1\le i<j\le N$ \cr
      \noalign{\smallskip}
      G(u) & if $1\le i\le N_1$ and $N_1+1\le j\le N$ \cr}
  \label{eq:12}
\end{equation}
where in (\ref{eq:12}), $N_1$ may take any value in the set
$\{1,2,\ldots,N-1\}$. Moreover, for any $N\ge3$ and $c^2=0$,
equation~(\ref{eq:8}) has one and only one odd, meromorphic solution, given by
(\ref{eq:11}). Both solutions~(\ref{eq:11}) and~(\ref{eq:12}) are well
known and describe either particles of the same type or of two different
types~\cite{19}.\par
%
%-------------------------------------------------------------------------
%
It should be noted that equation~(\ref{eq:8}) also has some singular solutions,
such as
\begin{equation}
  f_{ij}(u) = f_{ji}(u) = c\, \mbox{sgn}(u) = c\, [\theta(u) - \theta(-u)]
  \qquad 1\le i<j\le N \label{eq:13}
\end{equation}
where $\theta(u)$ denotes the Heaviside function.\par
%
%==========================================================================
%
\section{Solutions of Calogero-Sutherland-Moser models with exchange terms}
{}From a set of $N!$ functions $\F(x_1,\ldots,x_N)$, $\sigma \in
S_N$, satisfying equation~(\ref{eq:5}), one can construct in general $N!$
functions $\f{rs}(x_1,\ldots,x_N)$, defined by
\begin{equation}
  \f{rs} = \sum_{\sigma\in S_N} \V{rs}{\sigma} \F \label{eq:14}
\end{equation}
where $[f] \equiv \left[f_1 f_2 \ldots f_N\right]$ runs over all $N$-box
Young diagrams, $r$ and $s$ label the standard tableaux associated with
$[f]$, arranged in lexicographical order, and $\V{rs}{\sigma}$ denotes
Young's orthogonal matrix representation of $S_N$~\cite{20}. Such functions
$\f{rs}$ satisfy the system of equations
\begin{eqnarray}
  \partial_i \f{rs} & = & \sum_{j\ne i} f_{ij} \sum_t \f{rt} \V{ts}{p_{ij}}
     - c \sum_{j\ne i} \sum_t \biggl(\sum_{\sigma} \ts{ij} \V{rt}{\sigma} \F
     \biggr) \V{ts}{p_{ij}} \nonumber \\
  & & \mbox{} + \sum_{\sigma} \ls{i} \V{rs}{\sigma} \F \qquad
     i=1,2,\ldots,N. \label{eq:15}
\end{eqnarray}
In deriving~(\ref{eq:15}), use has been made of the first of the following
representation properties of $\V{rs}{\sigma}$,
\begin{equation}
  \V{rs}{\sigma \circ \sigma'} = \sum_t \V{rt}{\sigma} \V{ts}{\sigma'}
  \qquad \V{rs}{1} = \delta_{r,s} \label{eq:16}
\end{equation}
and of the first equality in (\ref{eq:4}).\par
%
%--------------------------------------------------------------------------
%
{}From~(\ref{eq:15}), it results that
\begin{eqnarray}
  \partial^2_{ii} \f{rs} & = & \sum_{j\ne i} (\partial_i f_{ij}) \sum_t
     \f{rt} \V{ts}{p_{ij}} + \sum_{j\ne i} f_{ij} \sum_t \left(\partial_i
     \f{rt}\right) \V{ts}{p_{ij}} \nonumber \\
  & & - c \sum_{j\ne i} \sum_t \biggl(\sum_{\sigma} \ts{ij} \V{rt}{\sigma}
     \partial_i \F\biggr) \V{ts}{p_{ij}} + \sum_{\sigma} \ls{i}
     \V{rs}{\sigma} \partial_i \F. \label{eq:17}
\end{eqnarray}
By using (\ref{eq:4}), (\ref{eq:5}), (\ref{eq:15}), and (\ref{eq:16})
again, and by summing over $i$, we obtain the following result for the
Laplacian of $\f{rs}$,
\begin{eqnarray}
  \Delta \f{rs} & = & \f{rs} \left(\sumij \bigl(f_{ij}^2 - c^2\bigr) +
     \sum_i \lambda_i^2\right) \nonumber \\
  & & \mbox{} + \sum_t \f{rt} \left(\sumij \bigl(\partial_i f_{ij}\bigr)
     \V{ts}{p_{ij}} + \sumijk f_{ij} f_{ik} \V{ts}{p_{ik}\circ
     p_{ij}}\right) \nonumber \\
  & & \mbox{} - c \sum_{\sigma} \sum_t \V{rt}{\sigma} \left(\sumijk \Bigl(
     f_{ij} \ts{ik} + f_{ik} \ts{kj}\Bigr) \V{ts}{p_{ik}\circ p_{ij}}
     \right) \F \nonumber \\
  & & \mbox{} + \sum_{\sigma} \sum_t \V{rt}{\sigma} \left(\sumij \bigl(
     \ls{i} + \ls{j}\bigr) f_{ij} \V{ts}{p_{ij}}\right) \F \nonumber \\
  & & \mbox{} + c^2 \sum_{\sigma} \sum_t \V{rt}{\sigma} \left(\sumijk
     \ts{ik} \ts{kj} \V{ts}{p_{ik}\circ p_{ij}}\right) \F \nonumber \\
  & & \mbox{} - c \sum_{\sigma} \sum_t \V{rt}{\sigma} \left(\sumij \bigl(
     \ls{i} + \ls{j}\bigr) \ts{ij} \V{ts}{p_{ij}}\right) \F. \label{eq:18}
\end{eqnarray}
We shall now proceed to evaluate the various terms on the right-hand side
of~(\ref{eq:18}).\par
%
%--------------------------------------------------------------------------
%
As
\begin{equation}
  p_{ik} \circ p_{ij} = p_{ij} \circ p_{jk} = p_{jk} \circ p_{ik}
  \label{eq:19}
\end{equation}
the last part of the second term becomes
\begin{eqnarray}
  \lefteqn{\sumijk f_{ij} f_{ik} \V{ts}{p_{ik}\circ p_{ij}}} \nonumber \\
  & = & \ordsum (f_{ij} f_{ik} + f_{jk} f_{ji} + f_{ki} f_{kj}) \Bigl(
      \V{ts}{p_{ik}\circ p_{ij}} + \V{ts}{p_{ij}\circ p_{ik}}\Bigr)
      \nonumber \\
  & = & c^2 \ordsum \Bigl(\V{ts}{p_{ik}\circ p_{ij}} + \V{ts}{p_{ij}\circ
      p_{ik}}\Bigr) \label{eq:20}
\end{eqnarray}
where in the last step we used the integrability conditions~(\ref{eq:6})
and~(\ref{eq:7}) of~(\ref{eq:5}). By applying~(\ref{eq:19}) again, the
summation over $i$, $j$,~$k$ in the third term on the right-hand side
of~(\ref{eq:18}) can be rewritten as
\begin{eqnarray}
  \lefteqn{\sumijk \Bigl(f_{ij} \ts{ik} + f_{ik} \ts{kj}\Bigr)
      \V{ts}{p_{ik}\circ p_{ij}}} \nonumber \\
  & = & \ordsum \biggl(\Bigl((f_{ij} + f_{ji}) \ts{ik} + (f_{jk} + f_{kj})
      \ts{ji} + (f_{ki} + f_{ik}) \ts{kj}\Bigr) \V{ts}{p_{ik}\circ p_{ij}}
      \nonumber \\
  & & \mbox{} + \Bigl((f_{ij} + f_{ji}) \ts{jk} + (f_{jk} + f_{kj})
      \ts{ki} + (f_{ki} + f_{ik}) \ts{ij}\Bigr) \V{ts}{p_{ij}\circ
      p_{ik}}\biggr) \label{eq:21}
\end{eqnarray}
and therefore vanishes owing to the antisymmetry of $f_{ij}$ in
$i$,~$j$, as shown in (\ref{eq:6}). The same is true for the
summations over $i$,~$j$ in the fourth and sixth terms as a consequence
of the antisymmetry of $f_{ij}$ and $\ts{ij}$, respectively. Finally,
by successively using~(\ref{eq:19}) and~(\ref{eq:4}), the summation over
$i$, $j$,~$k$ in the fifth term becomes
\begin{eqnarray}
  \lefteqn{\sumijk \ts{ik} \ts{kj} \V{ts}{p_{ik}\circ p_{ij}}}
      \nonumber \\
  & = & \ordsum \biggl(\Bigl(\ts{ji} \ts{ik} + \ts{kj} \ts{ji} + \ts{ik}
      \ts{kj}\Bigr) \V{ts}{p_{ik}\circ p_{ij}} \nonumber \\
  & & \mbox{} + \Bigl(\ts{ki} \ts{ij} + \ts{ij} \ts{jk} + \ts{jk}
      \ts{ki}\Bigr) \V{ts}{p_{ij}\circ p_{ik}}\biggr) \nonumber \\
  & = & - \ordsum \Bigl(\V{ts}{p_{ik}\circ p_{ij}} + \V{ts}{p_{ij}\circ
      p_{ik}}\Bigr). \label{eq:22}
\end{eqnarray}
\par
%
%--------------------------------------------------------------------------
%
By putting all results together, the Laplacian of $\f{rs}$ takes the simple
form
\begin{equation}
  \Delta \f{rs} = \left(\sumij \left(f_{ij}^2(x_i-x_j) + \left(\partial_i
  f_{ij}(x_i-x_j)\right) K_{ij} - c^2\right) + \sum_i \lambda_i^2\right)
  \f{rs} \label{eq:23}
\end{equation}
where $K_{ij} = K_{ji}$, $1\le i<j\le N$, are some operators, whose action
on $\f{rs}$ is defined by
\begin{equation}
  K_{ij} \f{rs} = \sum_t \f{rt} \V{ts}{p_{ij}}. \label{eq:24}
\end{equation}
Let us emphasize that equation~(\ref{eq:23}) is valid for any function $\f{rs}$
constructed from any solution of~(\ref{eq:5}) via
transformation~(\ref{eq:14}).\par
%
%---------------------------------------------------------------------------
%
In the special cases where $[f] = [N]$ or $\bigl[1^N\bigr]$, since
$V^{[N]}(p_{ij}) = - V^{[1^N]}(p_{ij}) = 1$, the operators $K_{ij}$ behave
as $I$ or $-I$, respectively. Hence $\varphi^{[N]} = \sum_{\sigma} \F$ and
$\varphi^{[1^N]} = \sum_{\sigma} (-1)^{\sigma} \F$, where $(-1)^{\sigma}$ is
the parity of permutation $\sigma$, are eigenfunctions of the operators $-
\Delta + \sum_{i\ne j} \left(f_{ij}^2 \pm \partial_i f_{ij} - c^2\right)$,
where the upper (resp.\ lower) sign corresponds to the former (resp.\
latter). For $f_{ij}$ given in~(\ref{eq:11}), these fit essentially the
Matsuo~\cite{16} and Cherednik~\cite{17} results.\par
%
%--------------------------------------------------------------------------
%
In the mixed symmetry cases where $[f]\ne [N]$, $\bigl[1^N\bigr]$, the
operators $K_{ij}$ have a nontrivial effect on the functions $\f{rs}$.
Provided the latter satisfy the conditions
\begin{eqnarray}
  \lefteqn{\f{rs}(x_1,\ldots,x_j,\ldots,x_i,\ldots,x_N)} \nonumber \\
  & = &\sum_t \f{rt}(x_1,\ldots,x_i,\ldots,x_j,\ldots,x_N) \V{ts}{p_{ij}}
      \qquad 1\le i<j\le N \label{eq:25}
\end{eqnarray}
which amount to
\begin{eqnarray}
  \lefteqn{\F(x_1,\ldots,x_j,\ldots,x_i,\ldots,x_N)} \nonumber \\
  & = & \Phi_{\sigma\circ p_{ij}}(x_1,\ldots,x_i,\ldots,x_j,\ldots,x_N)
      \qquad 1\le i<j\le N \label{eq:26}
\end{eqnarray}
for any $\sigma\in S_N$, the operators $K_{ij}$ may be interpreted as
permutation operators acting on the variables $x_i$ and $x_j$,
\begin{equation}
  K_{ij} x_j = x_i K_{ij} \qquad K_{ij} x_k = x_k K_{ij} \qquad k\ne i,j.
  \label{eq:27}
\end{equation}
\par
%
%---------------------------------------------------------------------------
%
It remains to examine under which conditions equation~(\ref{eq:5}) admits
solutions satisfying (\ref{eq:26}). This is readily done by
differentiating both sides of~(\ref{eq:26}) with respect to~$x_k$ and
using~(\ref{eq:5}) to calculate the derivatives. Equations~(\ref{eq:5})
and~(\ref{eq:26}) are found compatible if and only if all functions
$f_{ij}(u)$, $i\ne j$, coincide, hence in cases such as~(\ref{eq:11})
and~(\ref{eq:13}). For the former choice, equation~(\ref{eq:23}) becomes
\begin{equation}
  \left(- \Delta + \omega^2 \sumij \left(\csch \omega(x_i-x_j)\right)^2
  k(k-K_{ij}) + \sum_i \lambda_i^2\right) \f{rs} = 0 \label{eq:28}
\end{equation}
in the hyperbolic case ($c^2>0$); similar results are obtained in the
rational ($c^2=0$) and trigonometric ($c^2<0$) cases. Hence, we did prove
that from any solution of type (\ref{eq:3}),~(\ref{eq:26}) of the KZ
equations~(\ref{eq:1}), with $f_{ij}$ given in~(\ref{eq:11}), we can obtain
eigenfunctions $\f{rs}$ of the CSM Hamiltonians~\cite{1},
\cite{2},~\cite{3} with exchange terms~\cite{13}, which are characterized
by any given permutational symmetry $[f]$ under particle coordinate
exchange. To obtain wave functions describing an $N$-boson (resp.\
$N$-fermion) system, it only remains to combine $\f{rs}$ with a spin function
transforming under the same (resp.\ conjugate) irreducible
representation~$[f]$ (resp.~$[\tilde f]$) under exchange of the spin
variables. A similar result is valid for the Hamiltonian with
delta-function interactions~\cite{21}, corresponding to the functions
$f_{ij}$ given in~(\ref{eq:13}).\par
%
%--------------------------------------------------------------------------
%
As a last point, we would like to mention that when restricting ourselves
to solutions satisfying~(\ref{eq:26}) or
\begin{equation}
  K_{ij} \Phi = P^{(ij)} \Phi \qquad 1\le i<j\le N \label{eq:29}
\end{equation}
with $K_{ij}$ defined in~(\ref{eq:27}), equations~(\ref{eq:5}) and~(\ref{eq:1})
become equivalent to
\begin{equation}
  \partial_i \F = \left(\sum_{j\ne i} \left(f(x_i-x_j) + c\, \ts{ij}\right)
  K_{ij} + \ls{i}\right) \F \qquad i=1,2,\ldots,N \label{eq:30}
\end{equation}
and
\begin{equation}
  \partial_i \Phi = \left(\sum_{j\ne i} \left(f(x_i-x_j) + c\, \hat T^{(ij)}
  \right) K_{ij} + \lambda^{(i)}\right) \Phi \qquad i=1,2,\ldots,N
  \label{eq:31}
\end{equation}
respectively. In~(\ref{eq:31}), $\hat T^{(ij)}$ is an operator whose action
on functions~(\ref{eq:3}) is given by
\begin{equation}
  \hat T^{(ij)} \Phi = \sum_{\sigma} \ts{ij} \F e_{\sigma}. \label{eq:32}
\end{equation}
The corresponding operator $\hat T$ on $V \otimes V$ may be taken as
\begin{equation}
  \hat T = \sum_{k>l} \left(E_{kk}\otimes E_{ll} - E_{ll}\otimes E_{kk}
  \right). \label{eq:33}
\end{equation}
%
%=========================================================================
%
\section{Conclusion}
In the present paper, we did move one step further towards a deeper
understanding of the interplay between integrable systems and KZ~equations
(and, therefore, conformal models). We did indeed show that the Matsuo and
Cherednik results can be generalized to provide wave functions,
characterized by any given permutational symmetry, for some CSM models with
exchange terms, once solutions of the corresponding KZ~equations are known.
Such models include the spin generalizations of the original Calogero and
Sutherland models, as well as that with $\delta$-function interactions.\par
%
%---------------------------------------------------------------------------
%
Some interesting open questions are whether similar results may also hold
true for elliptic CSM models and for integrable models related to root
systems of algebras different from ${\cal A}_{N-1}$. The use of methods
similar to those employed in ref.~\cite{22} to construct generalizations of
Dunkl operators might prove to be helpful in finding proper answers.\par
%
%===========================================================================
%
\section*{Acknowledgment}
The author would like to thank T.~Brzezi\' nski for some helpful
discussions.\par
\vfill\eject
%
%===========================================================================
%


\begin{thebibliography}{99}

\bibitem{1} Calogero F 1969 {\sl J. Math. Phys.} {\bf 10} 2191, 2197; 1971 {\sl
J. Math. Phys.} {\bf 12} 419; 1975 {\sl Lett. Nuovo Cimento} {\bf 13} 411 \\
Calogero F, Ragnisco O and Marchioro C 1975 {\sl Lett. Nuovo Cimento} {\bf 13}
383

\bibitem{2} Sutherland B 1971 {\sl Phys. Rev.} A {\bf 4} 2019; 1972 {\sl Phys.
Rev.} A {\bf 5} 1372; 1975 {\sl Phys. Rev. Lett.} {\bf 34} 1083

\bibitem{3} Moser J 1975 {\sl Adv. Math.} {\bf 16} 1

\bibitem{4} Leinaas J M and Myrheim J 1988 {\sl Phys. Rev.} B {\bf 37} 9286 \\
Polychronakos A P 1989 {\sl Nucl. Phys.} B {\bf 324} 597; 1991 {\sl Phys.
Lett.}
{\bf 264B} 362 \\
Haldane F D M 1991 {\sl Phys. Rev. Lett.} {\bf 67} 937

\bibitem{5} Haldane F D M 1988 {\sl Phys. Rev. Lett.\/} {\bf 60} 635; 1991 {\sl
Phys. Rev. Lett.} {\bf 66} 1529 \\
Shastry B S 1988 {\sl Phys. Rev. Lett.\/} {\bf 60} 639

\bibitem{6} Chen H H, Lee Y C and Pereira N R 1979 {\sl Phys. Fluids\/}J{\bf
22}
187

\bibitem{7} Kazakov V A 1991 {\sl Random Surfaces and Quantum Gravity, Cargese
Lectures, 1990} ed O. Alvarez {\sl et al\/} (New York: Plenum)

\bibitem{8} Minahan J A and Polychronakos A P 1994 {\sl Phys. Lett.} {\bf 326B}
288

\bibitem{9} Olshanetsky M A and Perelomov A M 1983 {\sl Phys. Rep.} {\bf 94}
313

\bibitem{10} Dunkl C F 1989 {\sl Trans. Am. Math. Soc.} {\bf 311} 167

\bibitem{11} Polychronakos A P 1992 {\sl Phys. Rev. Lett.} {\bf 69} 703

\bibitem{12} Brink L, Hansson T H and Vasiliev M A 1992 {\sl Phys. Lett.}
{\bf 286B} 109

\bibitem{13} Ha Z N C and Haldane F D M 1992 {\sl Phys. Rev.} B {\bf 46} 9359
\\
Hikami K and Wadati M 1993 {\sl Phys. Lett.} {\bf 173A} 263 \\
Minahan J A and Polychronakos A P 1993 {\sl Phys. Lett.\/} {\bf 302B} 265 \\
Bernard D, Gaudin M, Haldane F D M and Pasquier V 1993 {\sl J. Phys. A: Math.
Gen.} {\bf 26} 5219

\bibitem{14} Brink L and Vasiliev M A 1993 {\sl Mod. Phys. Lett.} A {\bf 8}
3585

\bibitem{15} Knizhnik V G and Zamolodchikov A B 1984 {\sl Nucl. Phys.} B {\bf
247} 83

\bibitem{16} Matsuo A 1992 {\sl Invent. Math.\/}J{\bf 110} 95

\bibitem{17} Cherednik I V 1991 Integration of quantum many-body problems by
affine KZ~equations {\sl Preprint} Kyoto RIMS

\bibitem{18} Felder G and Veselov A P 1994 {\sl Commun. Math. Phys.\/}J{\bf
160}
259

\bibitem{19} Calogero F 1975 {\sl Lett. Nuovo Cimento} {\bf 13} 507

\bibitem{20} Rutherford D E 1948 {\sl Substitutional Analysis} (Edinburgh:
Edinburgh UP)

\bibitem{21} Lieb E H and Liniger W 1963 {\sl Phys. Rev.} {\bf 130} 1605 \\
Yang C N 1967 {\sl Phys. Rev. Lett.} {\bf 19} 1312; 1968 {\sl Phys.
Rev.\/}J{\bf 168} 1920

\bibitem{22} Buchstaber V M, Felder G and Veselov A P 1994 Elliptic Dunkl
operators, root systems, and functional equations {\sl Preprint} hep-th/9403178

\end {thebibliography}
\end{document}